\documentclass[aps,prl,reprint,groupedaddress,nofootinbib,nobibnotes,raggedbottom]{revtex4-2}
\usepackage[T1]{fontenc}
\usepackage{amsmath,amssymb,bm,microtype}
\usepackage{needspace}
\usepackage{graphicx}
\usepackage[colorlinks=true,linkcolor=blue,citecolor=blue,urlcolor=blue]{hyperref}
\newcommand{\dd}{\mathrm d}
\newcommand{\ii}{\mathrm i}

\newcommand{\op}{\mathrm{op}}

\newcommand{\tens}{\mathcal T}

\newcommand{\JU}{\mathcal J}

\begin{document}
\title{The Stringy Origin of Kerr Multipoles}
\author{M. Firrotta}
\email[Contact author: ]{mfirrotta@fudan.edu.cn}
\affiliation{Department of Physics and Center for Field Theory and Particle Physics,
Fudan University, Shanghai 200438, China}
\begin{abstract}
We construct a new family of exact open and closed bosonic string solutions in
flat spacetime that reproduce the root-Kerr and Kerr multipoles at every fixed multipole order in the large-mass and large-angular-momentum limit.
These solutions provide explicit classical data for identifying associated quantum string states and testing their correspondence
with rotating black holes.
\end{abstract}
\maketitle

The Kerr solution~\cite{Kerr} describes rotating black holes in general relativity
and provides a central benchmark for gravitational-wave tests of
gravity~\cite{GWTests}. Its mass and angular momentum fix the entire
hierarchy of permanent multipole moments~\cite{Hansen1974}, tightly constraining its
gravitational interactions~\cite{Israel:1970kp,Newman:1965tw,Bianchi:2024shc}. Explaining the microscopic origin of this
structure is a concrete challenge for quantum gravity. Scattering
amplitudes provide a precise formulation of the problem: the Kerr
three-point coupling encodes the full multipole hierarchy through its
exponential dependence on the specific spin~\cite{Guevara,AHO}.

Extending this description beyond three-point interactions is essential for
understanding Kerr dynamics. The gravitational Compton amplitude
describes graviton scattering off a rotating object and provides
building blocks for calculations of binary dynamics and radiation.
Its factorization poles are constrained by three-point amplitudes,
but additional contact interactions and dynamical response are not determined by the
permanent multipoles alone~\cite{ScheopnerVines}.

 Although all-spin proposals and
calculations from black-hole perturbation theory have substantially
advanced this problem~\cite{KerrCompton,KerrBHPT,KerrWorldline},
an independent microscopic derivation would provide a valuable
determination of the interactions beyond three points. String theory
offers a concrete route: amplitudes involving two arbitrary massive and spinning string
states and two gravitons are calculable~\cite{Firrotta:2024qel}, and their appropriate
classical limit can be compared directly with Kerr
scattering~\cite{StringCompton,SuperstringCompton}.

A necessary step is to identify string configurations with the
correct permanent multipoles. The leading Regge trajectory has a
well-defined classical limit, but its higher multipoles differ
from Kerr~\cite{Cangemi}, motivating the search for more general
string configurations and states~\cite{Searching}.

 In the present Letter,
we construct a new family of exact classical bosonic strings whose
normalized, period-averaged three-point interactions approach the electromagnetic
root-Kerr and gravitational Kerr couplings as the mass and angular momentum become large.

The construction starts from a connected endpoint trajectory on the
sphere with a prescribed measure. The polar
weights determine the electric multipoles, while the relative weights
between opposite circulations determine the magnetic multipoles.
A smooth finite-mass curve realizes both weights and determines
the entire worldsheet, satisfying the free-string equations,
boundary conditions, and Virasoro constraints. Its endpoint current, in the large-mass and large-angular momentum limit,
converges to root-Kerr, while the gravitational Kerr
coupling follows from the bulk stress tensor.

\textit{Exact strings from a generating curve---}
The classical solution of an open bosonic string with tension $\tens=(2\pi\ell_s^2)^{-1}$ and characteristic length $\ell_s$, obtained from the Polyakov action in conformal and temporal gauge, can be parametrized by the target time $t=L\tau$ and a smooth curve $\vec{r}(t)$ of period $T$ with unit speed
$|d\vec{r}/dt|=1$, and length scale $L=T/(2\pi)$.
 In particular, the embedding reads
\begin{equation}
 X^0=L\tau,\quad
 \vec{X}=\tfrac12\bigl[\vec{r}(L(\tau{+}\sigma))
                         {+}\vec{r}(L(\tau{-}\sigma))\bigr],
 \label{eq:embedding}
\end{equation}
with $0\leq\sigma\leq\pi$.
Its endpoints are located at $\vec{r}(t)$ and $\vec{r}(t+T/2)$. 
At fixed target time, every point of the string is the midpoint of two
points on the generating curve.

In conformal gauge, the classical Virasoro constraints are
$(\partial_\pm X)^2=0$, where
$\partial_\pm=\partial_\tau\pm\partial_\sigma$~\cite{Polchinski,GSW}.
Writing $t_\pm=L(\tau\pm\sigma)$, Eq.~\eqref{eq:embedding} gives
\begin{align}
 (\partial_\pm X)^2
 &=L^2\bigl(-1+|d\vec{r}/dt_{\pm}|^2\bigr)=0.
 \label{eq:virasoro}
\end{align}
Thus unit speed enforces both constraints, equivalently
$\partial_\tau X\cdot\partial_\sigma X=0$ and
$(\partial_\tau X)^2+(\partial_\sigma X)^2=0$.
The separated waves solve
$(\partial_\tau^2-\partial_\sigma^2)X^\mu=0$,
while periodicity gives the Neumann conditions
$\partial_\sigma X^\mu|_{\sigma=0,\pi}=0$.

%
In the rest frame, the charges representing mass and rotation are respectively
\begin{equation}
 M_\op=\frac{{\cal T}}{2}T,\quad
 \vec{a}:=\frac{\vec{J}_\op}{M_\op}
 =\frac1{2T}\int_0^{T}\dd t\,\vec{r}\times {d\vec{r}\over dt}.
 \label{eq:charges}
\end{equation}
A closed string follows by taking $0\leq\sigma<2\pi$ and allowing
two independent unit-speed curves of the same period in
Eq.~\eqref{eq:embedding}. Identical curves give a folded double cover,
with twice the open-string mass and angular momentum and the same
specific spin. Throughout this letter, dots denote derivatives with
respect to the target time $t$, and primes derivatives with
respect to the polar angle $\theta$.

\textit{A class of spiraling open string solutions---}
Let $R$ be the sphere radius, $b\in(0,\sqrt{2}{-}1)$ the dimensionless profile parameter and $8\lambda:=\Lambda/R$ a dimensionless parameter 
with the length $\Lambda\in \mathbb{R}_+$. A new family of open string solutions is provided by the following four-dimensional endpoint trajectory 
\begin{equation}
{\cal X}^{\mu}_{\lambda,b}(t(\theta))=(t_{\lambda,b}(\theta),\vec{r}_{\lambda,b}(t(\theta)))
\end{equation}
with spatial motion described by the coordinates 
\begin{equation}
\vec{r}_{\lambda,b}(t(\theta))=
 R(\sin\theta\cos\phi_{\lambda,b},\sin\theta\sin\phi_{\lambda,b},-\cos\theta),
 \label{eq:curve}
\end{equation}
and the target clock, which converts the angular curve parameter into physical time
\begin{equation}
t'_{\lambda,b}(\theta):=R\sqrt{1+\sin^2\theta\,  ({\phi'_{\lambda,b}}(\theta))^2}.
\label{clock}
\end{equation}
Since $t'_{\lambda,b}\geq R$, the clock has a smooth inverse; moreover
$|d{\vec{r}_{\lambda,b}/d\theta}|=dt_{\lambda,b}/d\theta$ proves unit speed exactly, satisfying Eq.~\eqref{eq:virasoro}.
Here $\theta$ denotes the polar angle measure along the $z$-direction from $-z$, while $\phi_{\lambda,b}(\theta)$ is the longitudinal angle, whose dependence on $\theta$ is specified by the parameters $\lambda$ and $b$ through
\begin{equation}
\phi'_{\lambda,b}(\theta):=\frac{\lambda^2\sin\theta}
 {\sqrt{1{+}\lambda^2\sin^2\theta}}
 \Big(f_b(-\cos\theta){+}\sin\theta\,G_b(-\cos\theta)\Big).
 \label{eq:phase}
\end{equation}
We choose the profile functions \(f_b\) and \(G_b\) to encode the multipole sequence of the string endpoint through the analytically continued spherical Poisson kernel \cite{Appell1887,LyndenBell2003}.
\begin{figure}[h!]
    \centering
    \includegraphics[width=7cm]{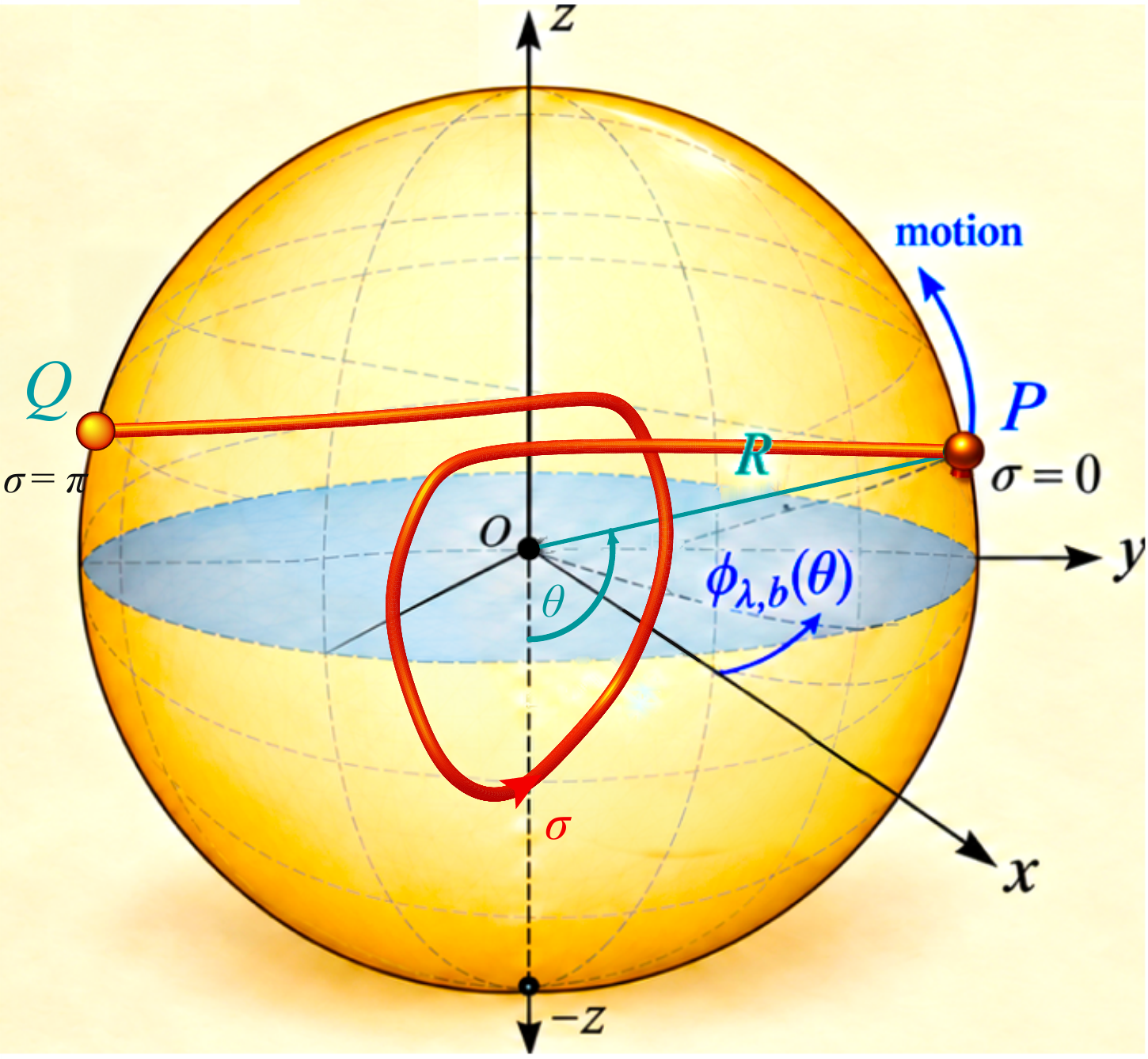}
    \caption{Instantaneous open-string configuration (red), with endpoints \(P\) and \(Q\) on a sphere of radius \(R\). The polar angle \(\theta\) is measured from \(-z\), and the longitudinal angle \(\phi_{\lambda,b}(\theta)\) from \(+x\) in the \(xy\) plane. Arrows indicate the endpoint motion (blue) and the \(\sigma\) profile (red).    }
\label{fig:setup}
\end{figure}
Their explicit form reads 
\begin{equation}
 f_b(u)=Re\frac{1{+}b^2}{d^{3/2}_{b,u}},\quad
 g_b(u)={1\over \rho_u} Im\frac{\ii b(3{-}b^2){-}u(1{-}3b^2)}{d_{b,u}^{3/2}},
 \label{app:weights}
\end{equation}
where $\rho_u:=\sqrt{1{-}u^2}$, $d_{b,u}:=1{-}b^2{-}2\ii bu$ and $G_b:=g_b/\rho_u$.

The primitives of the longitudinal angle and clock satisfy $t_{\lambda,b}(0)=\phi_{\lambda,b}(0)=0$ and their explicit forms are given in Eq.~\eqref{ph:global-phase-compact} and Eq.~\eqref{fullClock}, respectively. The coordinate frame of the endpoint solution is described in Fig.~\ref{fig:setup}.

\begin{figure}[h!]
    \centering
    \includegraphics[width=7cm]{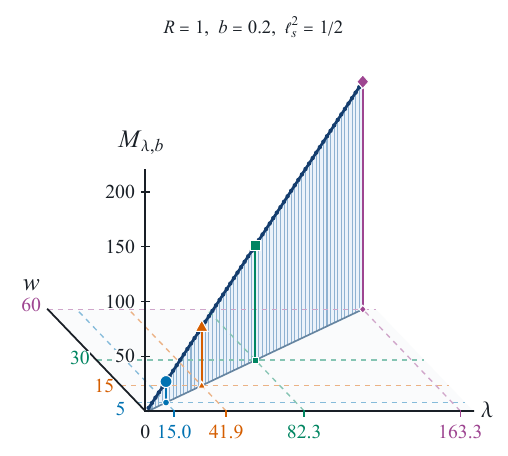}
    \caption{Mass \(M_{\lambda,b}\) of the open-string solutions for \(R{=}1\) and \(b{=}0.2\). The condition Eq.~\eqref{closure} fixes \(\lambda\) for each integer \(w\). Colored markers highlight \(w=5,15,30,60\). At large-\(\lambda\), the mass grows linearly with \(w\) .
    }
    \label{Fig1}
\end{figure}

\begin{figure}[h!]
    \centering
    \includegraphics[width=7cm]{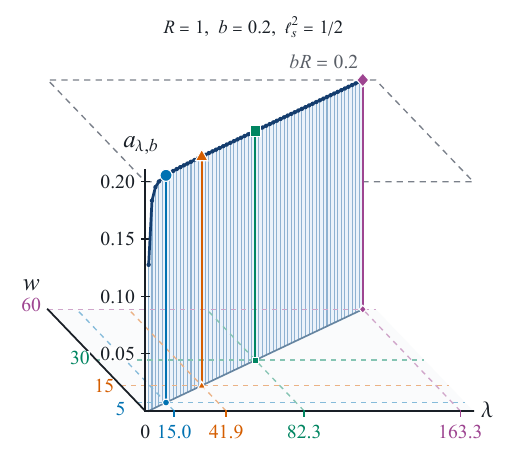}
    \caption{Angular length $a_{\lambda,b}$ for the same family of solutions  in Fig. \ref{Fig1} with \(R{=}1\) and \(b{=}0.2\). As \(\lambda\) increases, \(a_{\lambda,b}\) approaches \(bR{=}0.2\).
    }
    \label{Fig2}
\end{figure}

We are interested in solutions with periodicity $\theta{\in}[0,4\pi]$ and period $t_{\lambda,b}(4\pi){:=}T_{\lambda,b}$. After half a period we impose $\vec{r}_{\lambda,b}(t(\theta{+}2\pi)){=}diag(-1,-1,1) \vec{r}_{\lambda,b}(t(\theta))$, which translates into the condition
\begin{equation}\label{closure}
\phi_{\lambda,b}(2\pi)=(2w{+}1)\pi
\end{equation}
 with $w$ a non-negative integer. At fixed $R$ and $b$, the condition in Eq.~\eqref{closure} fixes the value of $\Lambda$ and therefore the possible values of $\lambda$ as a function of $w$. Each solution is identified by the parameters $R$, $b$ and $\lambda$ with the condition Eq.~\eqref{closure}. These parameters identify mass and rotation of the string. In Figs.\ref{Fig1}-\ref{Fig2} we show their behavior for a representative set of solutions.  

At large-$\lambda$, we find remarkably simple string solutions. In particular, for the longitudinal angle we obtain 
\begin{equation}
\phi_{\lambda,b}(\theta)\big|_{\lambda{\gg}1}=\lambda \phi_{\infty,b}(\theta) +O(1),
\end{equation}
while for the clock we have
\begin{equation}
t_{\lambda,b}(\theta)\big|_{\lambda{\gg}1}
=\lambda R t_{\infty,b}(\theta)
+O_b\!\left(R\log\lambda/\lambda\right),
\end{equation}
with explicit expressions given in Eq.~\eqref{phiExpL} and Eq.~\eqref{Cinf}, respectively.
In this limit, the condition Eq.~\eqref{closure} of the longitudinal angle simplifies, and the explicit solution is
\begin{equation}
\lambda+O_b\left(\log\lambda/\lambda\right)= {(2w{+}1)\pi\over 4H_b}
\end{equation}
where the function $H_b$ is defined in Eq.~\eqref{parmLlaphi}.
Mass and angular momentum also admit smooth expansions, which read 
\begin{align}
M_{\lambda,b}\big|_{\lambda{\gg}1}
&=\lambda\, M_{\infty,b}
+O_b\!\left(\mathcal T R\log\lambda/\lambda\right),\\
J_{\lambda,b}\big|_{\lambda{\gg}1}
&=\lambda\, J_{\infty,b} + O_b\!\left({\cal T}R^2/\lambda\right),
\end{align}
with $M_{\infty,b}:=4\mathcal T R$ and $J_{\infty,b}:=4\mathcal T bR^2$, leading to the following rotation length 
\begin{equation}
a_{\lambda,b}\big|_{\lambda{\gg}1}
=a_{\infty,b}+O_b\!\left(R\log\lambda/\lambda^2\right),
\end{equation}
with $a_{\infty,b}:=R b$. Remarkably, in the large-$\lambda$ limit, our construction yields string solutions with parametrically large mass and large angular momentum but finite rotation length $a_{\lambda,b}$, whose asymptotic value is determined by the fixed sphere radius $R$ and the parameter $b$ entering the spherical Poisson kernels in Eq.~\eqref{app:weights}.
In the same limit, the corresponding endpoint currents converge to the sources of root-Kerr and Kerr multipoles.
%

Moreover, the endpoint trajectory determines the full worldsheet through
a residual conformal reparametrization combined with a Weyl
rescaling. With $L=T_{\lambda,b}/(2\pi)$, we introduce
\begin{equation}
\theta_\pm:=2(\hat\tau\pm\hat\sigma),
\quad
L(\tau\pm\sigma)=t_{\lambda,b}(\theta_\pm),
\quad 0\leq\hat\sigma\leq\pi.
\end{equation}
The complete embedding is then
\begin{equation}
\widehat X^\mu_{\lambda,b}(\hat\tau,\hat\sigma)
=\frac12\left[
{\cal X}^\mu_{\lambda,b}\bigl(t_{\lambda,b}(\theta_+)\bigr)
+{\cal X}^\mu_{\lambda,b}\bigl(t_{\lambda,b}(\theta_-)\bigr)
\right].\label{OpEmbSol}
\end{equation}
Since $t'_{\lambda,b}>0$, this change of coordinates is
invertible. The worldsheet metric becomes
$h_{ab}=\Omega^2\eta_{ab}$, where
\begin{equation}
L^2\Omega^2=4t'_{\lambda,b}(\theta_+)
        t'_{\lambda,b}(\theta_-),
\quad
\hat h_{ab}:=\Omega^{-2}h_{ab}=\eta_{ab}.
\end{equation}
The separated dependence on $\hat\tau{\pm}\hat\sigma$ and
the null endpoint tangent ensure the wave equation and
Virasoro constraints. Extending the clock by
$t_{\lambda,b}(\theta{+}2\pi)
{=}t_{\lambda,b}(\theta){+}T_{\lambda,b}/2$
preserves both boundaries and their physical trajectories
$\vec r_{\lambda,b}(t)$ and
$\vec r_{\lambda,b}(t{+}T_{\lambda,b}/2)$.

A string profile at fixed target time is obtained by imposing
$\widehat X^0{=}[t_{\lambda,b}(\theta_+)
{+}t_{\lambda,b}(\theta_-)]/2{=}t$.

However, regularity of the clock reparametrization does not preclude
degeneracies of the induced metric. In the angular coordinates,
$
\gamma_{\hat\sigma\hat\sigma}
=2t'_{\lambda,b}(\theta_+)t'_{\lambda,b}(\theta_-)
\bigl(1-\dot{\vec r}_+\cdot\dot{\vec r}_-\bigr),
$
where $\dot{\vec r}_\pm:=
\dot{\vec r}_{\lambda,b}(t_{\lambda,b}(\theta_\pm))$
are the characteristic velocities in target time.
Since $t'_{\lambda,b}>0$, degeneracy occurs precisely when
these velocities coincide. This is required at the Neumann
endpoints and may also occur in the interior, giving rise to
folds or cusp events~\cite{Klimenko}. Such degeneracies belong
to the embedding and cannot be removed by a Weyl rescaling
of the auxiliary worldsheet metric.

\textit{From endpoint motion to root-Kerr---}
We now turn to the electromagnetic current generated by the
string endpoints. Following the notation in~\cite{Cangemi,Searching}, the normalized
current reads
\begin{equation}
\mathcal J^\mu_{\lambda,b}(k)
:=\frac1{T_{\lambda,b}}\int_0^{T_{\lambda,b}}
\dd t\,(1,\dot{\vec r}_{\lambda,b})^\mu
e^{\ii\vec k\cdot\vec r_{\lambda,b}},
\quad k^0=0.
\label{eq:root-current}
\end{equation}
The integration is over physical time and is independent of
the worldsheet parametrization. Periodicity gives $\JU^\mu(0){=}(1,\vec{0})$ and $k_\mu\JU^\mu{=}0$.

In the string rest frame $p^\mu{=}(M_{\lambda,b},\vec{0})$, we consider a photon with momentum $k{=}\kappa(0,\ii,0,1)$, polarization $\varepsilon_h{=}(1,0,h,0)$ and helicity $h{=}{\pm} 1$ such that $\vec{k}\times\vec{\varepsilon}_h{=}\ii h\vec{k}$.
Therefore, the three-point kinematics satisfy $k^2{=}0$ and $(p{+}k)^2{=}p^2$, hence $p{\cdot}k=0$. 
In this setup, the normalized three-point coupling ${\cal F}^{(h)}_{\lambda,b}:={\varepsilon_h{\cdot}\mathcal J_{\lambda,b}(k)/ \varepsilon_h{\cdot}\mathcal J_{\lambda,b}(0)}$  reads 
\begin{align}\label{3ptCo}
{\cal F}^{(h)}_{\lambda,b}=\frac1{T_{\lambda,b}}\int_0^{4\pi}\hspace{-1mm}d\theta\,{\cal W}_{\lambda,b}(\theta)\,e^{-\kappa R \sin\theta\cos\phi_{\lambda,b}-\ii\kappa R\cos\theta}
\end{align}
where we have introduced 
\begin{equation}\label{Bcoeff}
{\cal W}_{\lambda,b}(\theta):=t'_{\lambda,b}(\theta){-}h {r'}^{(y)}_{\lambda,b}(\theta),
\end{equation}
with the selected $y$-component of the angular tangent 
\begin{equation}\label{yVel}
{r'}^{(y)}_{\lambda,b}(\theta)=R\cos\theta\sin\phi_{\lambda,b}+R \phi'_{\lambda,b}\sin\theta\cos\phi_{\lambda,b}\,.
\end{equation}
The root-Kerr multipoles are obtained from the convergence of the large-$\lambda$ limit of the three-point coupling 
\begin{equation}
{\cal F}^{(h)}_{\sqrt{\textbf{Kerr}}}=\lim_{\lambda\rightarrow \infty}{\cal F}^{(h)}_{\lambda,b}=e^{h\kappa \,a_{\infty,b} }
\end{equation}
To establish this limit, we keep $R$ and $0<b<\sqrt2{-}1$
fixed and consider the sequence $\lambda$ selected
by Eq.~\eqref{closure}.
The clock estimates in Eqs.~\eqref{quiqui}--\eqref{prla:period-large-lambda}, together with the bounded term $R\cos\theta\sin\phi_{\lambda,b}$ in Eq.~\eqref{yVel}, leave the $\phi'_{\lambda,b}$-dependent contribution at leading order.
%
 Therefore, the three-point coupling Eq.~\eqref{3ptCo} in the large-$\lambda$ regime yields
\begin{align}
{\cal F}^{(h)}_{\lambda,b}\Big|_{\lambda\gg1}
={}&\frac1{8\lambda}\int_0^{4\pi}\dd\theta\,
\sin\theta\,\phi'_{\lambda,b}
(1-h\cos\phi_{\lambda,b})
\nonumber\\
&
e^{-\kappa R\sin\theta\cos\phi_{\lambda,b}
-\ii\kappa R\cos\theta}
+O_b(\lambda^{-1}).
\label{eq:current-leading}
\end{align}
The exact longitudinal angle is retained in the exponential,
since the bounded correction to
$\phi_{\lambda,b}=\lambda\phi_{\infty,b}+O_b(1)$
cannot be discarded.

The longitudinal dependence can now be expanded explicitly.
Writing $z:=\kappa R\sin\theta$ and using the modified Bessel functions $I_n(z):=\sum_{k=0}^{\infty}
\frac{1}{k!\,(k+n)!}
\left(\frac{z}{2}\right)^{2k+n}$,
we can employ the following expansion 
\begin{align}
&\,\qquad(1-h\cos\phi)\,e^{-z\cos\phi}
\nonumber\\
&=
\sum_{m\in\mathbb Z}(-)^m
\left[I_m(z){+}\frac h2
\bigl(I_{m-1}(z){+}I_{m+1}(z)\bigr)\right]
e^{\ii m\phi}.
\label{eq:longitudinal-Fourier}
\end{align}
From Eq.~\eqref{closure}, the odd Fourier modes cancel exactly. For $m\neq0$,
the identity
$\phi'_{\lambda,b}e^{\ii m\phi_{\lambda,b}}
=(\ii m)^{-1}\partial_\theta
e^{\ii m\phi_{\lambda,b}}$
allows integration by parts in
Eq.~\eqref{eq:current-leading} after the substitution of Eq.~\eqref{eq:longitudinal-Fourier}.
The boundary term vanishes because $\sin\theta=0$ at the
integration endpoints, while the remaining integral contributes
$O_b(\lambda^{-1})$. 

The Bessel coefficients and their $\theta$ derivatives are absolutely summable, uniformly in $\theta$ and for $\kappa$ in compact
subsets of $\mathbb C$. The zero-mode converges
locally uniformly by dominated convergence; Cauchy's formula then
permits extraction of every fixed Taylor coefficient. Thus, in the large-$\lambda$ limit, only the longitudinal zero-mode in Eq.~\eqref{eq:current-leading} survives.
Using
$t'_{\infty,b}=\sin\theta\,\phi'_{\infty,b}$,
the coupling becomes
\begin{align}
\lim_{\lambda\to\infty}{\cal F}^{(h)}_{\lambda,b}
={}&\frac18\int_0^{4\pi}\dd\theta\,
t'_{\infty,b}(\theta)\,
e^{-\ii\kappa R\cos\theta}
\nonumber\\
&
\left[I_0(\kappa R\sin\theta)
+hI_1(\kappa R\sin\theta)\right].
\end{align}
On each polar passage, see Fig.\eqref{fig:setup},
$t'_{\infty,b}
=|\sin\theta|\,[f_b(u)+\sin\theta\,G_b(u)]$,
with $u=-\cos\theta$. Combining the four passages and using
$\rho_u=\sqrt{1{-}u^2}$ and $g_b=\rho_u G_b$, we obtain
\begin{align}
\lim_{\lambda\to\infty}{\cal F}^{(h)}_{\lambda,b}
={}&\frac12\int_{-1}^{1}\dd u\,e^{\ii\kappa Ru}
\nonumber\\
&
\left[f_b(u)I_0(\kappa R\rho_u)
+h g_b(u)I_1(\kappa R\rho_u)\right].
\label{eq:current-zero-mode}
\end{align}
\begin{figure}[h!]
    \centering
    \includegraphics[width=8.5cm]{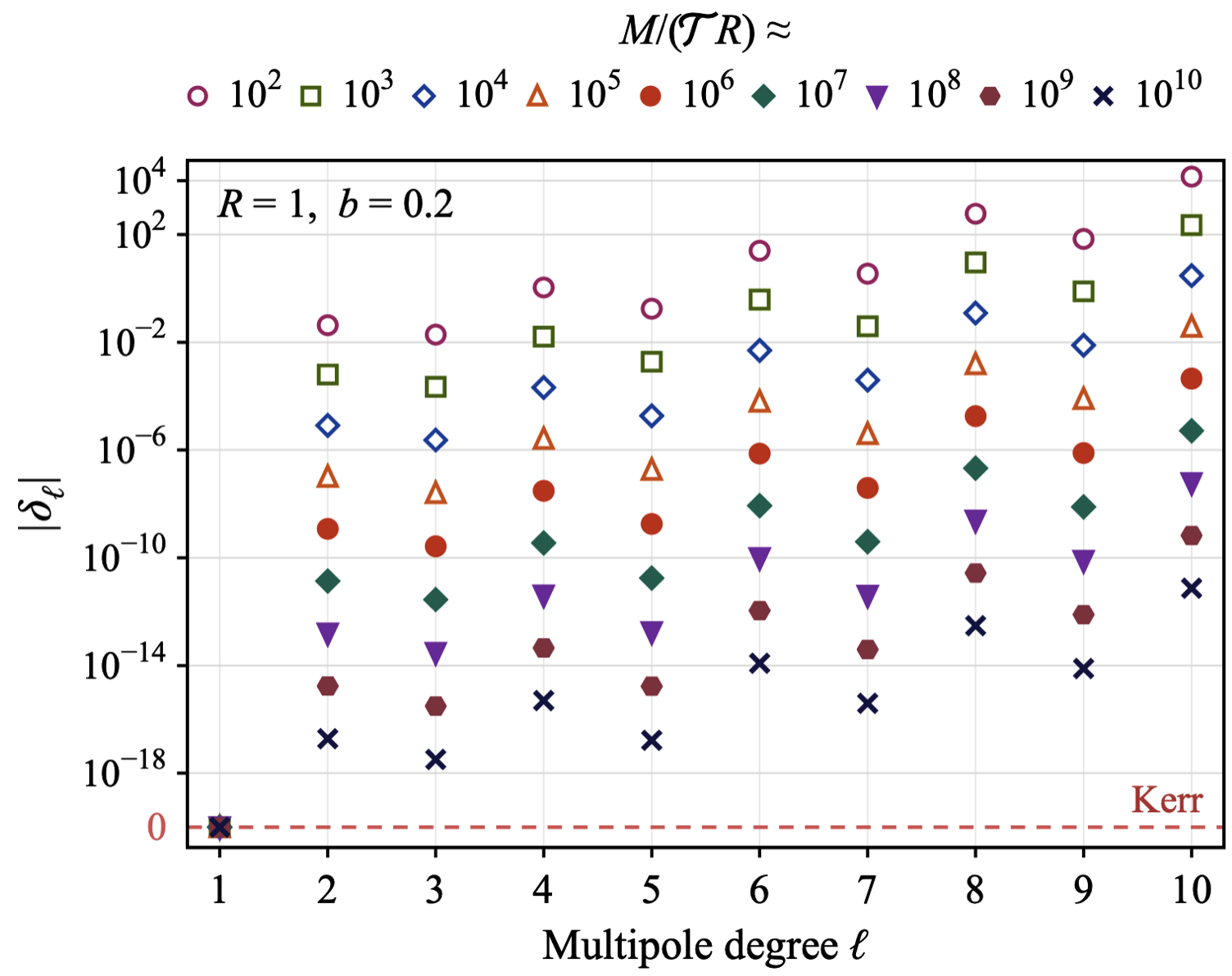}
\caption{Convergence of the first ten axial current multipoles to root Kerr.
Shown are the deviations $|\delta_\ell|{:=}|C_\ell{-}1|$ from Kerr multipoles, with $C_\ell{:=}(E_\ell{+}\mathrm{i}B_\ell)/(\mathrm{i}a_{\lambda,b})^\ell$, where $E_\ell$ and $B_\ell$ are electric and magnetic moments defined in Eqs.~\eqref{app:nonax-axial-electric}-\eqref{app:nonax-axial-magnetic}.
Each solution satisfies the condition Eq.~\eqref{closure} at $R{=}1$ and
$b{=}0.2$; legend values give rounded values of $M/(\mathcal T R)$. The vertical coordinate is
$\log_{10}(1{+}10^{20}|\delta_\ell|)$, with ticks labeled by
$|\delta_\ell|$. The dashed line denotes Kerr; all dipoles coincide
exactly at zero.}
\label{fig:multipoles}
\end{figure}
To evaluate Eq.~\eqref{eq:current-zero-mode}, we use the following expansion in Legendre polynomials $P_\ell(u)$
\begin{align}
&\,\qquad f_b(u)
=\sum_{\ell\geq0}(2\ell+1)
\operatorname{Re}(\ii b)^\ell P_\ell(u),\\
&g_b(u)=\rho_u G_b(u)
=\rho_u\sum_{\ell\geq1}\frac{2\ell+1}{\ell}
\operatorname{Im}(\ii b)^\ell P'_\ell(u).
\end{align}
with the required Bessel--Legendre integrals
\begin{align}
\frac12\int_{-1}^{1}\dd u\,
P_\ell(u)e^{\ii\kappa Ru}I_0(\kappa R\rho_u)
&=\frac{(\ii\kappa R)^\ell}{(2\ell+1)\ell!},\\
\frac12\int_{-1}^{1}\dd u\,
\rho_u P'_\ell(u)e^{\ii\kappa Ru}I_1(\kappa R\rho_u)
&=-\ii\ell\,
\frac{(\ii\kappa R)^\ell}{(2\ell+1)\ell!}.
\end{align}
Consequently, we can see that the coupling admits the multipole expansion
\begin{align}
\lim_{\lambda\to\infty}{\cal F}^{(h)}_{\lambda,b}
={}&\sum_{\ell=0}^{\infty}
\frac{(\ii\kappa R)^\ell}{\ell!}
\operatorname{Re}(\ii b)^\ell -\ii h\sum_{\ell=1}^{\infty}
\frac{(\ii\kappa R)^\ell}{\ell!}
\operatorname{Im}(\ii b)^\ell
\nonumber\\
={}&\cosh(\kappa Rb)+h\sinh(\kappa Rb)
=e^{h\kappa Rb}.
\end{align}
The even powers arise from the electric multipoles, while
the odd powers arise from the magnetic multipoles. In the
charge-normalized convention, their coefficients in Eqs.~\eqref{app:nonax-axial-electric}-\eqref{app:nonax-axial-magnetic} satisfy
\begin{equation}
\lim_{\lambda\rightarrow \infty}E_\ell+\ii B_\ell=(\ii a_{\infty,b})^\ell,
\quad
a_{\infty,b}=bR=\lim_{\lambda\to\infty}
\frac{J_{\lambda,b}}{M_{\lambda,b}},
\end{equation}
which identifies the root-Kerr
multipoles of the current on the complex null cone $k^2=0$ and $\vec{k}^2=0$
\begin{equation}
\mathcal J^0_{\infty,b}
=\cosh(a_{\infty,b}\kappa),
\quad
\vec{\mathcal J}_{\infty,b}
={\ii(\vec \epsilon_z{\times}\vec k)\over \kappa}\sinh(a_{\infty,b}\kappa).
\end{equation}
In Fig. \ref{fig:multipoles}, we show the convergence of the first ten axial current multipoles to their root-Kerr values as the mass increases, while in Fig. \ref{fig:nonaxisymmetric-degrees} we show the suppression of the non-axial multipoles Eqs.~\eqref{app:nonax-electric}-\eqref{app:nonax-magnetic}.

For each fixed nonzero $m$, the exact moments in
Eqs.~\eqref{app:nonax-electric} and~\eqref{app:nonax-magnetic} also vanish in this limit.
Use $0\le t'_{\lambda,b}-R\sin\theta\,\phi'_{\lambda,b}\le R$ and integrate
the $\phi'_{\lambda,b}e^{-\ii m\phi_{\lambda,b}}$ terms by parts.
The pole boundary terms vanish and the remaining angular factors are integrable.
Since $T_{\lambda,b}\sim8R\lambda$, this gives
$E_{\ell m},B_{\ell m}=O_{b,\ell,m}(R^\ell/\lambda)$ and
$\mathcal A_{X,\ell}=O_{b,\ell}(\lambda^{-1})$ at fixed $\ell$,
with $\mathcal A_{X,\ell}$ defined in Eq.~\eqref{app:nonax-degree}.

\textit{Sourcing Kerr from closed strings---} The generating-curve representation in Eq.~\eqref{eq:embedding},
extended to $0{\leq}\sigma{<}2\pi$ with identical characteristic
curves, defines a folded closed string. Its period-averaged
stress tensor factorizes into two generating-curve currents,
which can be evaluated directly from Eqs.~\eqref{eq:curve}-\eqref{clock}. This requires neither inversion of the
clock nor an explicit expression for $X^\mu(\tau,\sigma)$.
The doubled spatial domain gives
$M^{\mathrm{cl}}_{\lambda,b}=2M_{\lambda,b}$ and
$J^{\mathrm{cl}}_{\lambda,b}=2J_{\lambda,b}$,
leaving the specific spin unchanged.
The gravitational coupling follows from the period-averaged
bulk stress tensor, including the tension contribution,
\begin{align}
\Theta^{\mu\nu}_{\lambda,b}(k)
={}&\frac{\tens}{T_{\lambda,b}}
\int_0^{2\pi}\dd\tau\int_0^{2\pi}\dd\sigma\,
e^{\ii\vec k\cdot\vec X}
\nonumber\\
&\left(
\partial_\tau X^\mu\partial_\tau X^\nu
-\partial_\sigma X^\mu\partial_\sigma X^\nu
\right).
\end{align}
The characteristic coordinates $t_\pm=L(\tau\pm\sigma)$
factorize the average into two generating-curve currents \cite{Polchinski,DamourVilenkin2001},
giving the exact identity
\begin{equation}
\frac{\Theta^{\mu\nu}_{\lambda,b}(k)}
{M^{\mathrm{cl}}_{\lambda,b}}
=
\mathcal J^\mu_{\lambda,b}(k/2)
\mathcal J^\nu_{\lambda,b}(k/2),
\quad
M^{\mathrm{cl}}_{\lambda,b}=\tens T_{\lambda,b}.
\end{equation}
The half-momentum arguments follow from the midpoint embedding.
Contracting with the graviton polarization
$\varepsilon_h^\mu\varepsilon_h^\nu$ therefore yields
\begin{equation}
{\cal G}^{(2h)}_{\lambda,b}(\kappa)
:=
\frac{\varepsilon_{h\mu}\varepsilon_{h\nu}
\Theta^{\mu\nu}_{\lambda,b}(k)}
{\varepsilon_{h\mu}\varepsilon_{h\nu}
\Theta^{\mu\nu}_{\lambda,b}(0)}
=
\left({\cal F}^{(h)}_{\lambda,b}(\kappa/2)\right)^2.
\end{equation}
Using the root-Kerr limit established above, we obtain
\begin{equation}
{\cal G}^{(2h)}_{\textbf{Kerr}}(\kappa):=\lim_{\lambda\to\infty}{\cal G}^{(2h)}_{\lambda,b}(\kappa)
=e^{h\kappa a_{\infty,b}}.
\end{equation}
The gravitational multipoles on the chosen complex null ray, satisfy
\begin{align}
\lim_{\lambda\to\infty}
\frac{\Theta^{00}_{\lambda,b}
{+}\sum_i\Theta^{ii}_{\lambda,b}}
{M^{\mathrm{cl}}_{\lambda,b}}
=\cosh(\kappa a_{\infty,b}),\\
\lim_{\lambda\to\infty}
\frac{-2\Theta^{0y}_{\lambda,b}}
{M^{\mathrm{cl}}_{\lambda,b}}
=\sinh(\kappa a_{\infty,b}).
\end{align}
Thus the mass and current multipoles of the linearized exterior
gravitational field satisfy, at each fixed $\ell$,
\begin{equation}
\lim_{\lambda\to\infty}
\frac{M_{\ell;\lambda,b}+\ii S_{\ell;\lambda,b}}
{M^{\mathrm{cl}}_{\lambda,b}}
=(\ii a_{\infty,b})^\ell,
\quad a_{\infty,b}=Rb,
\end{equation}
reproducing the Kerr multipole sequence \cite{Thorne}.

\textit{Conclusion and outlook---}
In the present Letter, we have constructed a class of exact bosonic
string solutions in flat spacetime whose three-point couplings
reproduce the root-Kerr and Kerr multipoles in the large-mass and large-angular-momentum limit. A similar class of solutions also exists in superstring theory.

 The characteristic multipole structure of Kerr is thus realized through
the geometry and motion of an extended rotating string. 

These explicit string
profiles provide the classical data needed to identify the associated
coherent string states~\cite{BianchiFirrotta,Searching} and investigate
their quantum corrections and higher-point
interactions~\cite{Firrotta:2024qel,StringCompton}.

Beyond three points, early recursive constructions
of opposite-helicity gravitational Compton amplitudes encounter
spurious poles above quantum spin two. Although pole-free completions
are available, factorization alone does not fix their contact
interactions~\cite{AoudeHaddadHelset2022}. Identifying the physical
vertex operators associated with our solutions would provide a
string-theoretic top-down determination of these interactions, including
transitions invisible to the diagonal three-point coupling~\cite{ScheopnerVines}.

A decisive next step is therefore to compute their Compton amplitudes
and compare them with Kerr results from quantum field theory and
black-hole perturbation
theory~\cite{KerrCompton,KerrBHPT,KerrWorldline,KerrCompton2026}.

Such a comparison would determine whether the agreement extends
beyond permanent multipoles. In particular, the periodic string
motion underlying the stationary current may leave
observable signatures in tidal response and absorption. The static
conservative response~\cite{IvanovZhou} and its dynamical and
dissipative extensions~\cite{SakethZhouIvanov,GoldbergerLiRothstein}
provide distinct benchmarks for establishing how far the
correspondence extends from the multipole structure of Kerr
to its dynamics.

Moreover, another important step is to establish whether the string configurations remain long-lived as their multipoles approach
Kerr~\cite{IengoRusso,SklirosCopelandSaffin}. Following their evolution under radiation and gravitational backreaction~\cite{DamourVeneziano} would determine whether the multipole agreement survives the onset of self-gravity. Therefore, these solutions provide a concrete setting for testing whether
the agreement of stationary multipoles extends to a dynamical correspondence between strings and rotating black holes~\cite{HorowitzPolchinski,CeplakEtAl}.


\textit{Acknowledgments---}
I would like to thank M. Bianchi, R. Emparan, P. Di Vecchia, E. Kiritsis and A. Tseytlin for useful comments and discussions.

\clearpage
\section*{End Matter}
\allowdisplaybreaks[4]

\setcounter{equation}{0}
\renewcommand{\theequation}{A\arabic{equation}}
\renewcommand{\theHequation}{A\arabic{equation}}

\textit{Appendix---}

\textit{Exact longitudinal angle.}
The primitive of Eq.~\eqref{eq:phase} uses the Lauricella function
\cite{Karlsson,Lauricella2020}
\begin{align}
 &F_D^{(n)}(\alpha;\boldsymbol\beta;\gamma;\mathbf z)
 =\sum_{\mathbf m\in \mathbb{Z}^n_{\geq0}}
 \frac{(\alpha)_{m_1+\cdots+m_n}}{(\gamma)_{m_1+\cdots+m_n}}
 \prod_{j=1}^{n}\frac{(\beta_j)_{m_j}}{m_j!}z_j^{m_j}\nonumber\\*
 &=\frac{\Gamma(\gamma)}{\Gamma(\alpha)\Gamma(\gamma{-}\alpha)}
 \int_0^1dy\, y^{\alpha-1}(1{-}y)^{\gamma-\alpha-1}
 \prod_{j=1}^{n}(1{-}z_jy)^{-\beta_j}
 \label{ph:FD-series}
\end{align}
where $(c)_m=\Gamma(c+m)/\Gamma(c)$ and
$\boldsymbol\beta,\mathbf z,\mathbf m$ have $n$ components.
The series converges for $|z_j|<1$; the Euler integral requires
$\Re\gamma>\Re\alpha>0$, with branches continued from $\mathbf z=0$.
On $j\pi\leq\theta\leq(j+1)\pi$, $j=0,1,2,3$,
\begin{equation}
 \phi_{\lambda,b}(\theta)=
 \mathsf {\cal A}_{\lambda,b}(\theta)+(-)^j\mathsf {\cal B}_{\lambda,b}(\theta)
 +(j{+}\mathbb{P}_j)\mathsf {\cal B}_{\lambda,b}(\pi),
 \label{ph:global-phase-compact}
\end{equation}
with $\mathbb P_j=(1-(-1)^j)/2$ and
\begin{align}
 &\mathcal A_{\lambda,b}(\theta)
 :=2\lambda^2(1+b^2)
 Re\left\{\frac{\sin^2(\theta/2)}{(1+\mathrm i b)^3}\right.\nonumber\\*[-1mm]
 &\qquad\,\,\,\,  F_D^{(3)}\!\Bigl(
 1;\tfrac12,\tfrac12,\tfrac32;2;
 r_\lambda^+(\theta),\left.
 r_\lambda^-(\theta),\zeta_b(\theta)
 \Bigr)\right\},
 \label{prla:A}\\
 &\mathcal B_{\lambda,b}(\theta):=\lambda^2Im\left[
 \mathcal Y_{\lambda,b}^{(1/2)}(\theta)
 -\chi_b\mathcal Y_{\lambda,b}^{(3/2)}(\theta)\right],
 \label{prla:B}
 \\
 &\, \qquad\qquad\qquad\mathcal Y_{\lambda,b}^{(\alpha)}(\theta)
 :=\frac{(\sin^{2}(\theta/2))^\alpha}{\alpha}\nonumber\\*
 & F_D^{(4)}\!\Bigl(
 \alpha;\tfrac12,\tfrac12,\tfrac12,\tfrac32;\alpha{+}1;\sin^2\tfrac\theta2,
 r_\lambda^+(\theta),
 r_\lambda^-(\theta),
 \zeta_b(\theta)\Bigr),
 \label{prla:Y}\\
 &r^{\pm}_\lambda(\theta):=\left(-2\lambda^2\pm2\lambda\sqrt{1+\lambda^2}\right)\sin^2{\theta\over 2},\\
 &\zeta_b(\theta):=\frac{4\ii b}{(1+\ii b)^2}\sin^2{\theta\over 2},\quad 
 \chi_b:=\frac{2(1-3b^2)}{(1+\ii b)^3}.
 \label{ph:arguments}
\end{align}

For fixed $0<b<1$, uniformly on $0\leq\theta\leq4\pi$,
\begin{align}\label{phiExpL}
&\phi_{\lambda,b}(\theta)\big|_{\lambda{\gg}1}
=\lambda\phi_{\infty,b}(\theta)+O_b(1),\\
&\phi_{\infty,b}(\theta)
=\mathcal A_{\infty,b}(\theta)
{+}(-1)^j\mathcal B_{\infty,b}(\theta)
{+}2(j{+}\mathbb P_j)H_b,
\\
&\mathcal A_{\infty,b}(\theta)
=2(1{+}b^2)\sqrt{x}\,
Re\!\left[
\frac{
F_D^{(2)}\!\left(
\tfrac12;\tfrac12,\tfrac32;\tfrac32;
x,\zeta_b(\theta)\right)}
{(1+\ii b)^3}
\right],\\
&\mathcal B_{\infty,b}(\theta)
=H_b-\arctan\frac{b\cos\theta}{S_b(\theta)}
-\frac{2b\cos\theta}{S_b(\theta)D_b(\theta)},
\\\label{parmLlaphi}
&D_b(\theta):=\sqrt{(1{-}b^2)^2{+}4b^2\cos^2\theta},\,\,\,S^2_b(\theta):=\frac{D_b(\theta){+}1{-}b^2}{2}\nonumber\\*
&\,\qquad\quad H_b:=\arctan b+\frac{2b}{1+b^2}
\end{align}
Here $x=\sin^2(\theta/2)$, $S_b>0$, and the branch inherited from
the exact primitive has $\sqrt{x}=|\sin(\theta/2)|$,
$\mathcal A_{\infty,b}(0)=\mathcal B_{\infty,b}(0)=0$ and
$\mathcal B_{\infty,b}(\pi)=2H_b$.

\newpage
\textit{Exact clock.}
For Eq.~\eqref{clock}, define
\begin{align}
 x_b(\theta)
 &:=
 \frac{2b|\sin\theta|}
 {\delta_b(\theta)},\quad
 \nu_b:=\frac{\lambda(1{+}b^2)}{2b}\\
 &\delta_b(\theta):=1{+}b^2{+}\sqrt{(1{-}b^2)^2{+}4b^2\cos^2\theta}
 \label{prla:clockaux}
\end{align}
and let $h_{\lambda,b,k}^{(s)}$, $k=1,\ldots,12$, $s=\pm1$, be the roots of
\begin{align}
 &0=(h{-}s)^6(h{+}s)^2
 \bigl[(h^2{+}1)^2{+}4\nu_b^2h^2\bigr]
 \nonumber\\*
 &+\frac{\lambda^4}{b^4}h^2(h^2{-}b^2)(h^2{+}1)
 \bigl[(1{-}3b^2)h{+}s(1{+}b^2)\bigr]^2.
 \label{prla:clockroots}\\
 &\boldsymbol\beta_{\rm clk}
 =
 \Big(
 1,1,2,
 \underbrace{\tfrac12,\ldots,\tfrac12}_{8},
 \underbrace{-\tfrac12,\ldots,-\tfrac12}_{12}
 \Big),
 \nonumber\\*
 &\boldsymbol\eta_{\lambda,b}^{(s)}
 =
 \Bigl(
 \pm\mathrm i,s,\pm b^{-1},\pm b,
 \pm\mathrm i\bigl[\sqrt{1+\nu_b^2}+\nu_b\bigr],
 \nonumber\\*
 &\qquad
 \pm\mathrm i\bigl[\sqrt{1+\nu_b^2}-\nu_b\bigr],
 h_{\lambda,b,1}^{(s)},\ldots,h_{\lambda,b,12}^{(s)}
 \Bigr).
 \label{prla:clockvectors}
\end{align}
Each $\pm$ denotes the ordered pair $(+,-)$. The quarter-clock is
\begin{align}
 \mathcal C_{\lambda,b}^{(s)}(\theta)
 ={}&\frac{1{+}b^2}{b}\,x_b(\theta) F_D^{(23)}\!\left(
 1;\boldsymbol\beta_{\rm clk};2;
 x_b(\theta)\boldsymbol\eta_{\lambda,b}^{(s)}
 \right).
 \label{prla:C}
\end{align}
Continue the Euler radicals from $+1$ at $x_b=0$ to $x_b=b$
to select the real branch. Then $x_b(0)=x_b(\pi)=0$,
$x_b(\pi/2)=b$ and $\mathcal C_{\lambda,b}^{(s)}(0)
=\mathcal C_{\lambda,b}^{(s)}(\pi)=0$.
Set $c_\pm:=\mathcal C_{\lambda,b}^{(\pm)}(\pi/2)$ and
$\sigma_j=(-1)^{j(j-1)/2}$. On
$j\pi/2\leq\theta\leq(j+1)\pi/2$, $j=0,1,2,3$,
\begin{equation}\label{fullClock}
t_{\lambda,b}(\theta)
=R\Bigl[
(j-\mathbb P_j)c_+
+2\mathbb P_j c_{\sigma_j}
+(-1)^j\mathcal C_{\lambda,b}^{(\sigma_j)}(\theta)
\Bigr].
\end{equation}

Extend by $t_{\lambda,b}(\theta+2\pi)
=t_{\lambda,b}(\theta)+2R(c_++c_-)$ for $0\leq\theta\leq2\pi$; hence
\begin{equation}
T_{\lambda,b}=t_{\lambda,b}(4\pi)=4R(c_++c_-).
\end{equation}
For $0<b<1$, the signs of $f_b\pm g_b$ are those of
$1-3b^2\pm(1+b^2)x_b$. Since $0\leq x_b\leq b$, nonnegativity requires
\begin{align}
 1-3b^2-(1+b^2)b
 &=(1+b)(1-2b-b^2)\geq0,
\end{align}
Thus $0<b\leq b_c:=\sqrt2-1$; the main text uses the strict
range $b<b_c$. At $b_c$, $f_b-g_b$ vanishes at $u=0$, but
\begin{equation}\label{quiqui}
 t'_{\lambda,b}(\theta)
 =R\sqrt{1+\sin^2\theta\,(\phi'_{\lambda,b}(\theta))^2}
 \geq R>0.
\end{equation}
For fixed $R$ and $0<b<b_c$, as $\lambda\to\infty$,
\begin{align}
&t_{\lambda,b}(\theta)\big|_{\lambda{\gg}1}
=\lambda R  t_{\infty,b}(\theta)
+O_b\!\left(R\log\lambda/\lambda\right),\\\label{Cinf}
&{\cal C}_{\infty,b}^{(s)}(\theta)
:={}1-\frac{1+s x_b}{1-s x_b}
\sqrt{\frac{b^2-x_b^2}{b^2(1+x_b^2)}}
\nonumber\\*
&+2s x_b F_D^{(3)}\!\left(
\tfrac12;\tfrac32,\tfrac12,-1;\tfrac32;
-x_b^2,\frac{x_b^2}{b^2},
\frac{(1{-}b^2)x_b^2}{2b^2}
\right),
\end{align}
where $0\leq\theta\leq\pi/2$, $x_b=x_b(\theta)$ and
$c_{\infty,\pm}:=\mathcal C_{\infty,b}^{(\pm)}(\pi/2)$.
Replacing $\mathcal C,c$ by $\mathcal C_\infty,c_\infty$ in the
bracket of Eq.~\eqref{fullClock} defines the dimensionless $t_{\infty,b}$,
with $t_{\infty,b}(\theta+2\pi)=t_{\infty,b}(\theta)+4$.
The odd-$s$ terms cancel at $x_b=b$, giving
\begin{equation}
T_{\lambda,b}\big|_{\lambda{\gg}1}
=8R\lambda
+O_b\!\left(R\log\lambda/\lambda\right).
\label{prla:period-large-lambda}
\end{equation}

\textit{Mass and angular momentum.}
For the open string, Eq.~\eqref{eq:charges} gives $J_x=J_y=0$ and
\begin{align}
&\,\qquad M_{\lambda,b}
=\frac{\mathcal T}{2}T_{\lambda,b}
=2\mathcal T R(c_++c_-),\\
J_{\lambda,b}
&=4\mathcal T R^2\lambda^2\,Im\Bigl[
\mathcal Y_{\lambda,b}^{(3/2)}(\pi)
{-}(1{+}\chi_b)\mathcal Y_{\lambda,b}^{(5/2)}(\pi)
\nonumber\\*
&\hspace{36mm}
+\chi_b\mathcal Y_{\lambda,b}^{(7/2)}(\pi)
\Bigr],
\end{align}
Here $J_{\lambda,b}=J_z$. At fixed $R$, $\mathcal T$ and $0<b<b_c$,
\begin{align}
M_{\lambda,b}\big|_{\lambda{\gg}1}
&=4\mathcal T R\lambda
+O_b\!\left(\mathcal T R\log\lambda/\lambda\right),\\
J_{\lambda,b}\big|_{\lambda{\gg}1}
&=4\mathcal T bR^2\lambda
-\mathcal T R^2H_b/\lambda\nonumber\\*
&\quad+O_b\!\left(\mathcal T R^2\log\lambda/\lambda^3\right),\\
a_{\lambda,b}:=\frac{J_{\lambda,b}}{M_{\lambda,b}}
&=bR+O_b\!\left(R\log\lambda/\lambda^2\right).
\end{align}
The folded closed string has twice these charges and the same $a_{\lambda,b}$.

\textit{Nonaxisymmetric multipoles.}
Expand the exact current,
 Eq.~\eqref{eq:root-current}, in
$P_\ell^m(u)e^{\mathrm im\phi_{\lambda,b}}$, with argument $u{=}{-}\cos\theta$ and
$P_\ell^m{=}(-)^m(1{-}u^2)^{m/2}d^mP_\ell/du^m$ with normalization
$N_{\ell m}{=}\sqrt{(\ell{-}m)!/(\ell{+}m)!}$.
Half-period symmetry removes odd $m$. For $m=0,2,\ldots\le\ell$,
\begin{align}
 E_{\ell m}&=\frac{N_{\ell m}R^\ell}{T_{\lambda,b}}
 \int_0^{4\pi}d\theta\,t'_{\lambda,b}P_\ell^m(u)
 e^{-\mathrm im\phi_{\lambda,b}},
 \label{app:nonax-electric}\\
 B_{\ell m}&=\frac{N_{\ell m}R^{\ell+1}}{(\ell+1)T_{\lambda,b}}
 \int_0^{4\pi}d\theta\,e^{-\mathrm im\phi_{\lambda,b}}
 \nonumber\\*[-1mm]
 &\quad\left[
 \sin^2\theta\,\phi'_{\lambda,b}\frac{dP_\ell^m(u)}{du}
 +\frac{\mathrm im}{\sin\theta}P_\ell^m(u)\right].
 \label{app:nonax-magnetic}
\end{align}
Retain the signed $\sin\theta$ over the full period; the $m$-term
is regular at the poles for even $m>0$ and absent at $m=0$.
The axial moments $E_{\ell0}=E_\ell$, $B_{\ell0}=B_\ell$ are
\begin{align}
 E_\ell&=\frac{R^\ell}{T_{\lambda,b}}
 \int_0^{4\pi}d\theta\,t'_{\lambda,b}P_\ell(u),
 \label{app:nonax-axial-electric}\\
 B_\ell&=\frac{R^{\ell+1}}{(\ell+1)T_{\lambda,b}}
 \int_0^{4\pi}d\theta\,\sin^2\theta\,\phi'_{\lambda,b}
 \frac{dP_\ell(u)}{du},
 \label{app:nonax-axial-magnetic}
\end{align}
Thus $E_{00}=1$, $B_{10}=a_{\lambda,b}=J_{\lambda,b}/M_{\lambda,b}$ and
$E_{\ell,-m}=(-1)^mE_{\ell m}^{*}$ (similarly for $B$).
At fixed $R,b$, the closure condition selects $\lambda=\lambda_w$.
Figure~\ref{fig:nonaxisymmetric-degrees} shows the separate norms
\begin{equation}
 \mathcal A_{X,\ell}^{\,2}
 =\frac{2}{a_{\lambda,b}^{2\ell}}
 \sum_{\substack{2\le m\le\ell\\m\ {\rm even}}}|X_{\ell m}|^2,
 \qquad X=E,B,
 \label{app:nonax-degree}
\end{equation}
evaluated with the exact finite clock, angle and physical $a_{\lambda,b}$.
Both sums vanish for $\ell=1$.

\onecolumngrid
\par
\vspace{12pt}
\setbox0=\vbox\bgroup
\noindent\begin{minipage}{\textwidth}
 \centering
 \includegraphics[width=\textwidth]{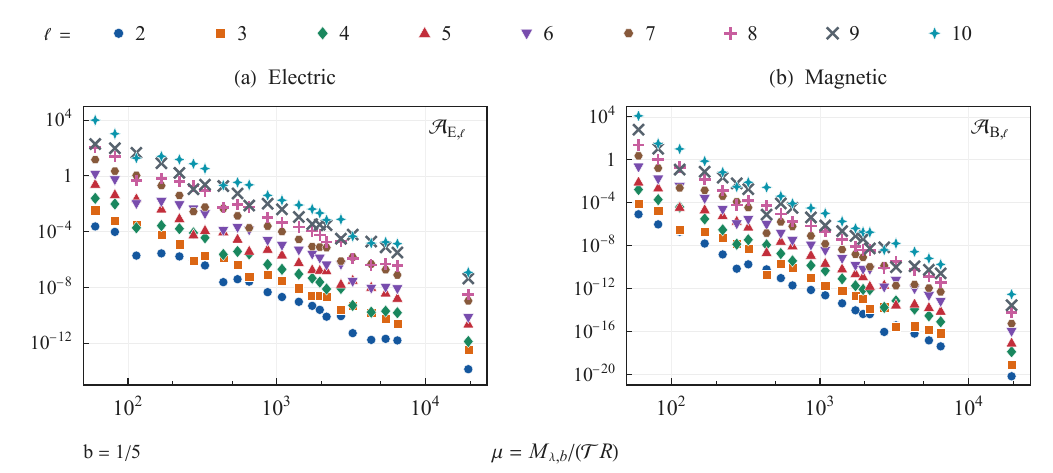}
 \makeatletter\def\@captype{figure}\makeatother
 \caption{Nonaxisymmetric electric (a) and magnetic (b) norms,
 Eq.~\eqref{app:nonax-degree}, for $\ell=2,\ldots,10$ at fixed $R$ and
 $b=1/5$. Points denote 22 exact-closure branches constrained by Eq.~\eqref{closure}, $5\leq w\leq1800$,
 normalized by the finite physical $a_{\lambda,b}$.
 Both axes are logarithmic, with different vertical ranges.
 The norms approach their vanishing root-Kerr values as
 $M_{\lambda,b}/(\mathcal T R)$ increases.}
 \label{fig:nonaxisymmetric-degrees}
\end{minipage}
\egroup
\box0
\end{document}